\documentclass{elsart}
\usepackage{amssymb}
\usepackage{graphicx}

\def \beq{\begin{equation}}
\def \eeq{\end{equation}}
\def \beqar{\begin{eqnarray}}
\def \eeqar{\end{eqnarray}}

\begin{document}
\begin{frontmatter}
\title{The Minority Game with interactions}

\author{In\'es Caridi\thanksref{mail}},
\author{Horacio Ceva\thanksref{mail2}}
\address{Departamento de F{\'{\i}}sica, Comisi{\'o}n Nacional 
de Energ{\'\i }a At{\'o}mica, Avda. del Libertador 8250, 1429 Buenos 
Aires, Argentina}
\thanks[mail]{email: caridi@cnea.gov.ar}
\thanks[mail2]{email: ceva@cnea.gov.ar}
\begin{abstract}
We partially modify the rules of the Minority Game (MG) by introducing 
some degree of local information in the game, which is only available 
for some agents, called the \emph{interacting} agents. Our work shows 
that, for small values of the new parameter of the model (the 
\emph{fraction of interacting agents}), there is an improvement of the 
use of the resources with respect to the MG, while as this number grows 
the response of the system changes, and ends up behaving worst than the 
usual MG.  
\end{abstract}
\end{frontmatter}

\section{Introduction}

The minority game (MG) is an adaptive game suitable to study competitive 
systems whose available resources are finite \cite{Challet}-\cite 
{econofisica}. In spite of the simplicity of the rules of the game, it 
presents a very rich behavior and interesting properties.

In this game there is a number of players (agents) that must choose, 
\emph{ independently}, one of two alternatives at each time step, 
splitting into two groups. The winners are those who stay at the 
minority side in that step. As we explain in the next section, there are 
three main ingredients in the game:\ (i) a \emph{public} history, with 
the information of the \emph{m} previous minority sides;\ (ii) some 
prescriptions (strategies) to play with, and (iii) a reward system, 
whereby points are given to winning agents and strategies. These 
ingredients induce an emergent coordination whereby the system can be 
more efficient (in a sense explained below) than a random game. In fact, 
this is the main reason for the interest shown by many people on the 
behavior of the model.

A recent line of research introduces modifications into the MG, designed 
to approach the behavior of the players to that of the agents in a real 
market. Agents in an economy take decisions based on heterogeneous 
external information available at each time, plus internal information, 
including their personal preferences. The heterogeneous external 
information can be either global, such as an aggregate variable, or 
local. Some of those investigations study the effect of changes in the 
external information available in the MG. The work of M. Paczuski et al. 
\cite{Maya Paczuski} develops a version of Kauffman's model with certain 
rules of the MG. Other works introduce versions of the MG with local 
information. Between these, that of Kalinowski et al. \cite{Kalinowski} 
places the agents on a circle;\ every agent gets the previous decisions 
of her $(m-1)$ neighbors (half to her right, half to her left) as an 
input information; together with her own decision, she builds up a 
history of \emph{m} bits. In this form there is a \emph{local} history 
for each player. The rest of the rules are as in the usual MG. \ S. 
Moelbert et al. \cite{Moelbert LocalMG} recently introduced a 
\emph{local MG}, where they use a local determination of the minority 
and the corresponding local histories.

In this work some kind of local information will be made available to a 
fraction $p$ of the $N$\ agents ($0\leq p\leq 1)$; we will refer to them 
as the \emph{interacting agents.} As we will see, they will be able to 
change their original decisions, in order to be in the minority of their 
neighborhoods.

In the next section we describe the model, and present the results 
found;\ in section \ref{conclusiones} some conclusions are drawn.

\section{The MG with interactions}

\subsection{The model \label{descripcion del MGint}}
 In the MG model there are $N$ agents, each one with $s$ strategies 
 distributed at random (with reposition) at the beginning of the game. 
 The strategies assign one output (0 or 1) for every one of the $2^{m}$ 
 possible inputs ($i.e$., histories). As a result of their choices, at 
 each step of the game the players split into two groups of size $N_{0}$ 
 and $N_{1},$ such that $N_{0}+N_{1}=N$. Every winner agent gets a point 
 after each step of the game. Besides, out of all the strategies playing 
 in the game, those that correctly predict the minority side of the time 
 step $t$, obtain a reward, called a virtual point. The agents use the 
 virtual points to establish the performance of her strategies and to 
 choose, at each step of the game, the more successful one to play with.

The main variable considered in this model is $\sigma $, the standard 
deviation of the difference $(N_{1}-N_{0})$

\begin{equation}
\sigma ^{2}=\frac{1}{T}\sum_{t=1}^{T}(N_{1}-N_{0})^{2}\label{dispersion}
\end{equation}

where \emph{T} is the number of time steps of the game. This variable 
measures the form in which the resources of all the players are used. 
When $ \mid N_{1}-N_{0}\mid $ is small, the minority group is bigger, 
and the agents, as a whole, receive more points. Hence, smaller values 
of $\sigma $ indicate a better use of the resources by the population. 
As mentioned above, the interest on this model is mainly due to the fact 
that for certain values of the variables \emph{m, s, }and\emph{\ N,} \ 
the standard deviation turns out to be smaller than in the case of a 
random game. The characterization of this collective behavior generated 
many studies of the MG, with a variety of techniques: numerical 
simulations \cite{Challet}, \cite {Manuca}; mean-field approximations 
\cite{Savit}; equivalence with spin-glass models \cite{Challet SpinG}; 
thermal treatments \cite{Cavagna TMG}; etc.

To introduce local information, we need to arrange the agents in such a 
form that it makes sense to talk about the neighborhood of every one of 
them. Hence, we distribute all the players on a square lattice with 
periodic boundary conditions. The positions of the $pN$ interacting 
agents on the lattice are chosen randomly, at the beginning of the game. 
At each time step the agents follow the usual rules of the MG, but the 
interacting agents $ (I.A.)$ are given the extra opportunity to modify 
their bets, after knowing what their nearest neighbors $I.A.$ will do in 
the $same$ step. As they were distributed randomly,\ \ every one of them 
can have between zero and four $ I.A.$ as nearest neighbors. This agent 
will try to be in the minority of the group formed by her $I.A.$ nearest 
neighbors, plus herself. Hence, if more than half of her neighbors 
choose one side, then the $I.A.$ will chose the other, regardless of 
what her best strategy says; but if half of the neighbors choose each 
side, then the $I.A.$ will decide in agreement with the prescription of 
her best strategy, $i.e.$ the one with more virtual points. With this 
rule, she can eventually choose one option that none of her strategies 
prescribe for a given history. Once all the agents have made their 
choice, they make their moves \emph{simultaneously}. Finally, one 
obtains the minority as usual, by considering the actions of all the 
players. As far as the players are concerned, points are distributed 
with the same rule than in the MG, but whenever a player wins because 
she plays against the majority of her local neighborhood, then no points 
are added to her strategies.

The fact that the $I.A.$ positions are chosen randomly raises the 
question of \ the relevance of the actual number of links between 
interacting agents, ${\ell ,}$ for every specific realization ($i.e.$ 
for every sample), for a fixed value $pN.$ We considered this factor, 
and found that in this case it only has minor effects \cite{Caridi}.

\subsection{Results \label{resultados}}

We have made extensive numerical simulations, finding $\sigma ^{2}/N$ as 
a function of both $p$ and $2^{m}/N$. In the following we will refer to 
$\mu _{r}\equiv \sigma ^{2}/N$ as the\emph{\ reduced variance}, and use 
$z\equiv 2^{m}/N,$ for short. As is well known \cite{Manuca}, the graph 
of $\mu_r$ \emph{vs} $z$ displays two different informational behaviors: 
an efficient phase for $z < z_c$, and an inefficient phase for bigger 
values of $z$. Qualitatively, $z_c$ is the greatest value of $z$ for 
which the system is dominated by a period-two, maladaptive dynamics that 
is explained below. 

\begin{figure}[tbp]
\includegraphics[width=12cm,clip]{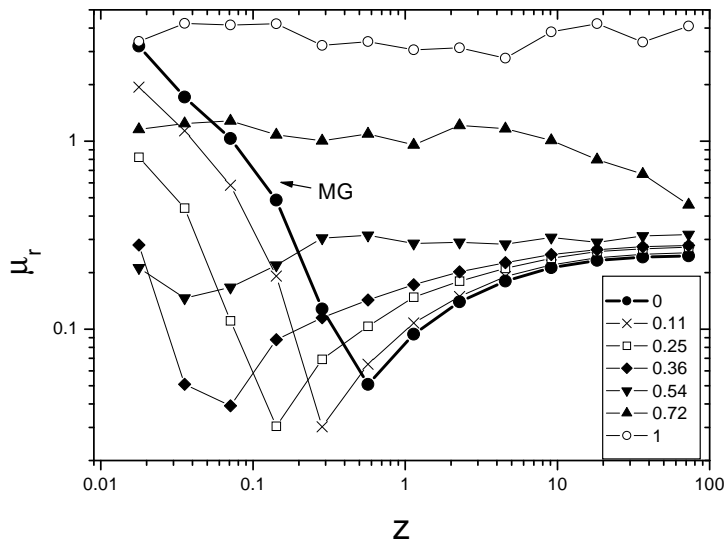}
\caption{Reduced variance $\mu_r$ as a function of \emph{z}, for $N$=225 
agents, and several values of $p$, indicated in the inset. Out of a 
total of $10^5$ time steps, data was collected in the last $9 \times 
10^4$ steps.}
\label{disper7}
\end{figure}

In Fig.\ref{disper7} we can see the ensuing changes in the reduced 
variance, as $p$ increases from $p=0$, $i.e.$ the MG with no interacting 
agents, to $p=1$, when all the agents are interacting. For small values 
of $p$ the function $\mu _{r}(z)$ has a shape similar to that of the MG 
case, but there is a shift to the left of its minimum, implying an 
improved efficiency for small values of $m$. Moreover, the minimum $\mu 
_{min}$ is smaller than the corresponding MG value. As $p$ increases, the 
shape of the curve changes, and for $p=1$ the reduced variance is \ 
roughly constant, and the efficiency is clearly worst than in the MG 
case.

In the following we will restrict ourselves to consider the first, 
informationally efficient region. We now will relate this behavior with 
the crowd - anticrowd picture \cite{Manuca}, \cite{Johnson}.

The minority game has, for $m=2$, a remarkable period-two dynamics 
(PTD), whereby during the odd appearances of a given history, the 
resulting minority is random, but then in the following, even 
appearance, the minority is just the opposite to the previous one. 
Moreover, if we distinguish even and odd occurrences of the histories, 
\emph{i.e.} we calculate the standard deviation $\sigma_{e} 
(\sigma_{o})$ for the time steps in which the histories occur an even 
(odd) number of times, it turns out that $\sigma _{o}$ is of the order of 
$\sqrt{N}/4$, while $\sigma _{e}$ $\gg \sigma _{o}.$\ If $m=2,$ this is 
true for almost all the steps of the game; as $m$ increases, this rule 
is true only in a (decreasing) number of steps, and when $z\approx 10$ 
or greater, the opposite happens:\ the `even' minority tends to be the 
same than the previous, \ `odd' one. This was first observed and 
explained by Manuca \emph{ et.al.} \cite{Manuca}, as due to the 
formation of crowds (see also \cite {Johnson}, \cite{Caridi2}). In a few 
words, this happens as follows: during the first (odd) occurrence of a 
history, the strategies of the players have the same amount of virtual 
points, and therefore they will pick side randomly, so that 
$N_{0}\approx N_{1}$; thereafter, the winner strategies will get a 
point. During the second (even) appearance of the same history, agents 
having those strategies will bias the outcome, choosing the same side as 
before. Hence, the minority will be just the opposite than the previous 
one; on the other hand, $N_{0}$ and $N_{1}$ will be very different, so 
that $ \sigma _{e}$ $\gg \sigma _{o}.$ Now the strategies receiving 
virtual points are those complementary to the first set; in this form 
all strategies have roughly the same amount of points, closing a 
`cycle'.

We begin by studying the effect of the interactions on the $m=2$ case; 
as we will see, for small values of $p$ the system still follows the 
PTD, but as $ p $ increases the period-two regime will disappear.

\begin{figure}[tbp]
\begin{tabular}{ccc}
\includegraphics[width=4.5cm,clip]{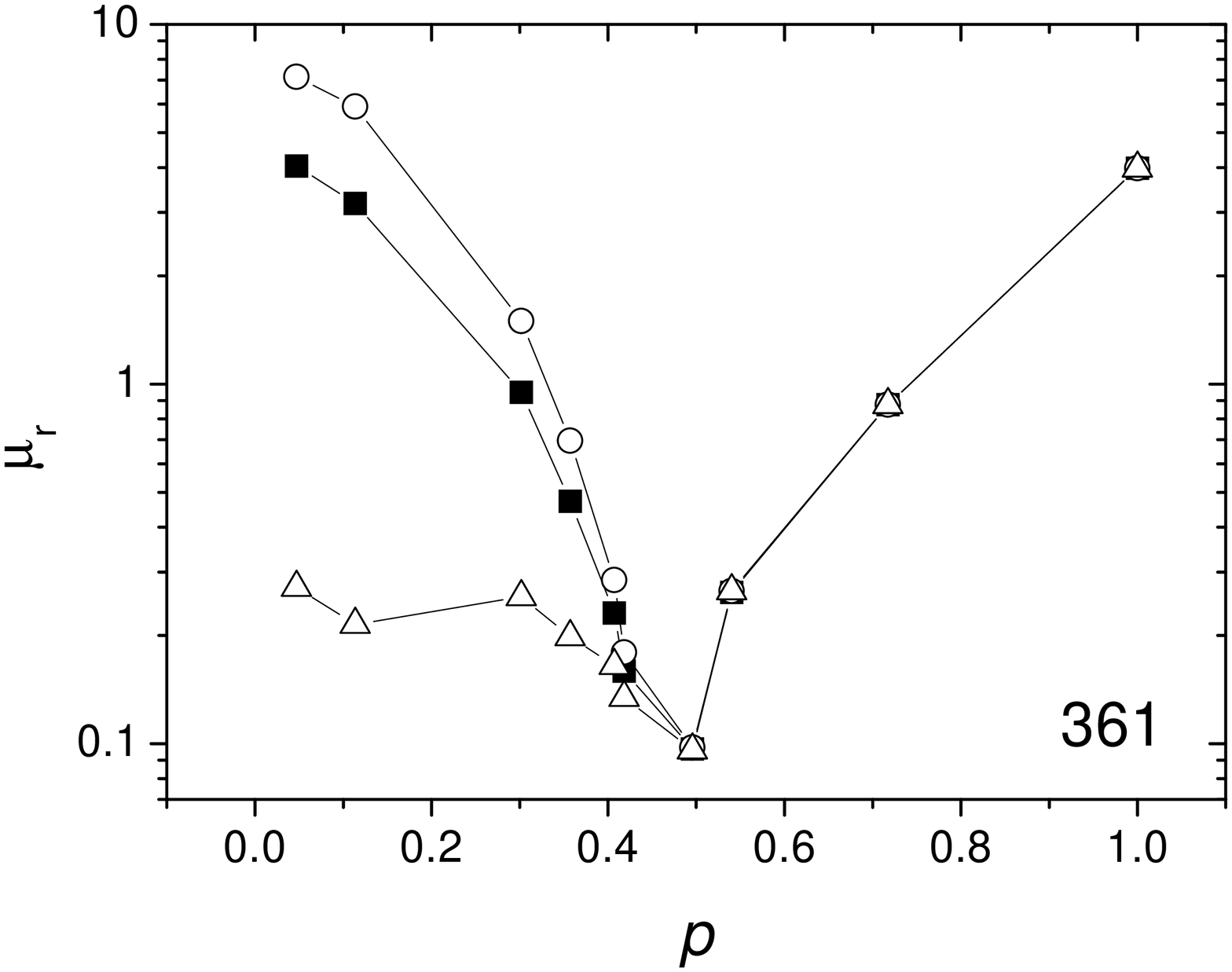}  & \includegraphics
[width=4.5cm,clip]{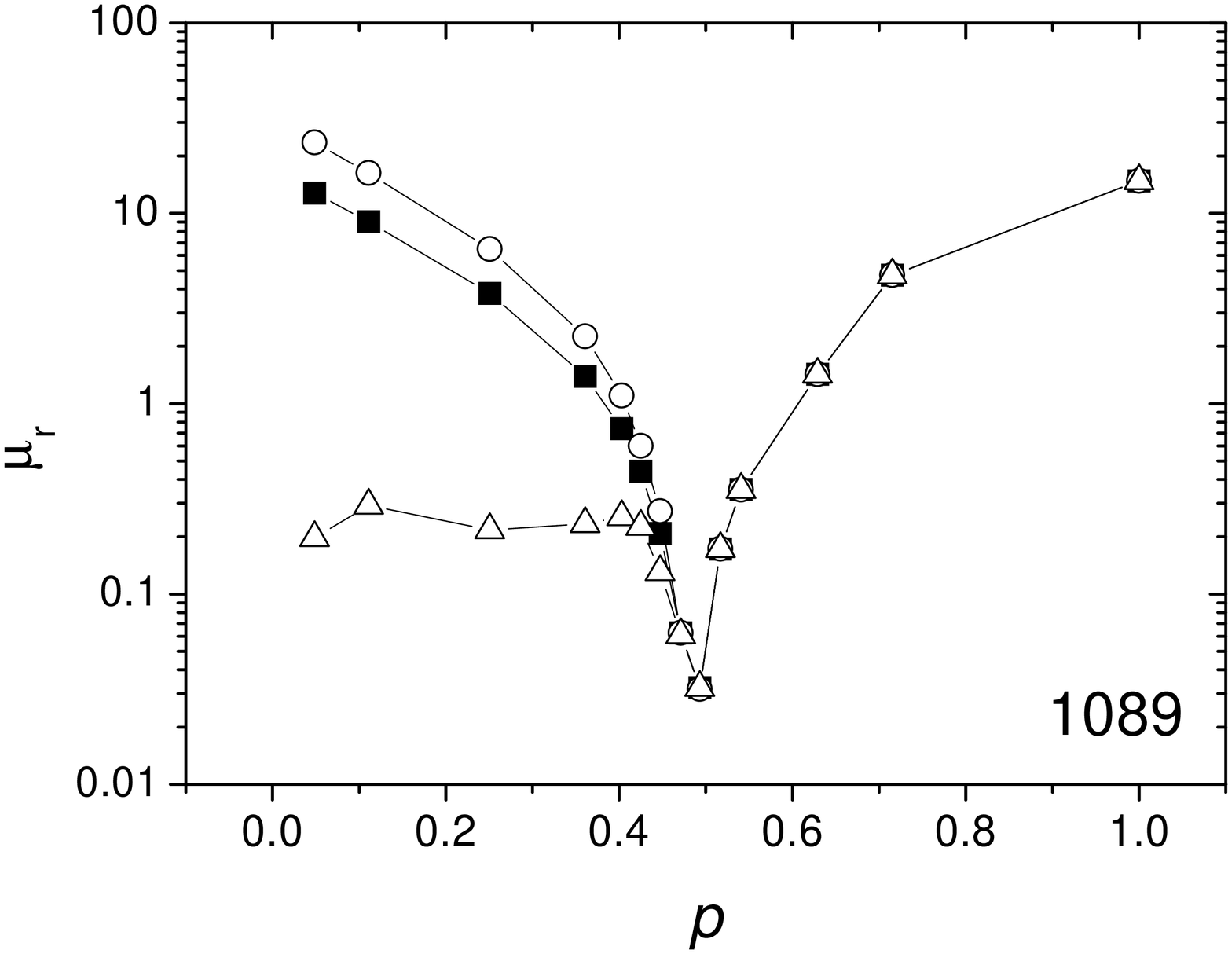} &
\includegraphics[width=4.5cm,clip]{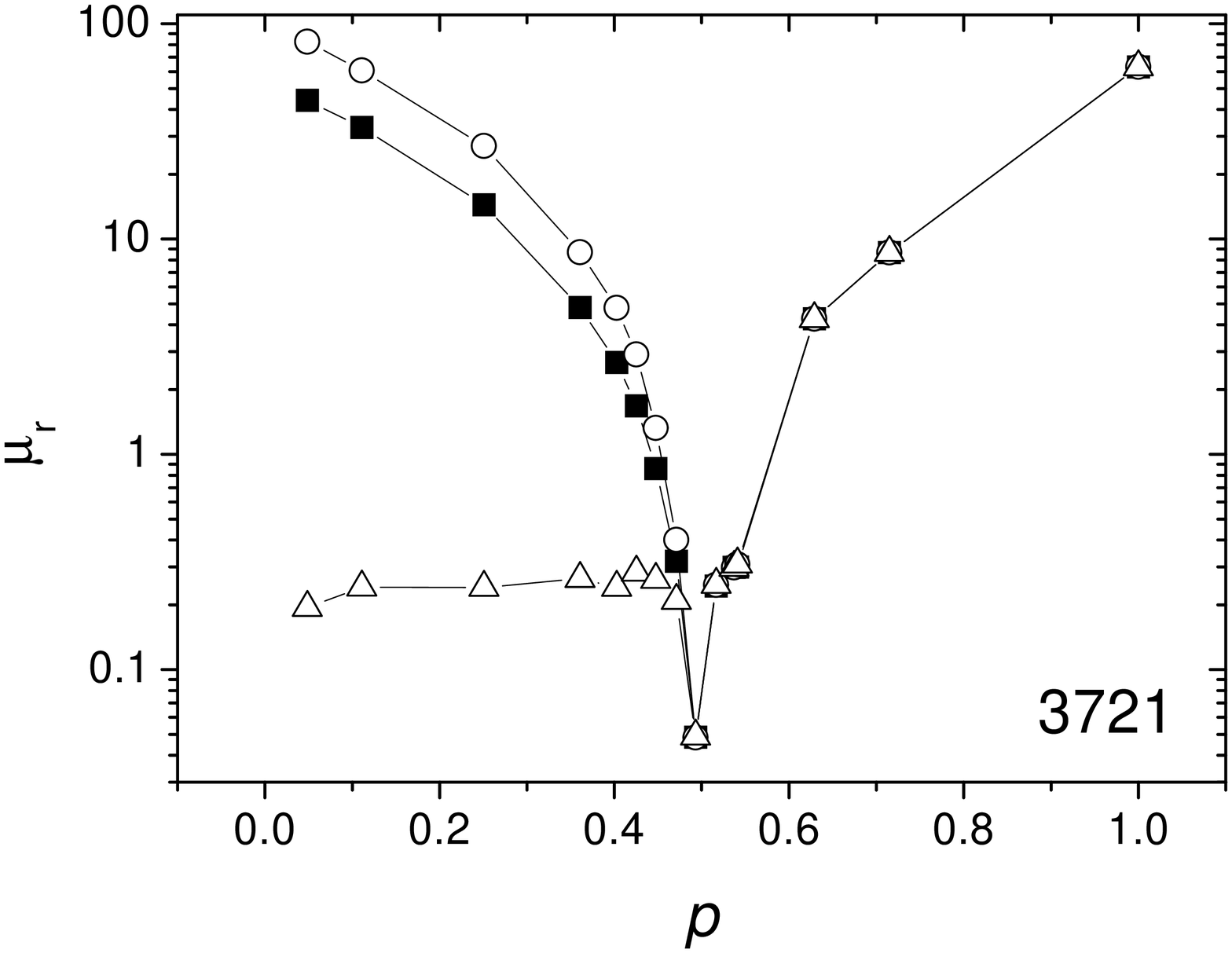} \\
\end{tabular}
\caption{Reduced variance $\mu_r$ as a function of $p$, for $m=2$. The 
different values of the reduced variance are calculated using $\sigma_e$ 
(empty circles), the whole dispersion $\sigma$ (filled squares), and 
$\sigma_o$ (empty triangles).}
\label{disper4}
\end{figure}

Figure \ref{disper4} shows data for $\mu _{r}$ as a function of $p,$ for 
$ N $ =361, 1089 and 3721 (these are just three representative examples 
of all the cases we have considered). We discarded the first  10$^{4}$ 
time steps for $N=361$, and 5 10$^{4}$ for bigger values of $N$, and 
collected data along the next 10$^{5}$ time steps. All the results shown 
are averages over 32 independent realizations.
 
Each set has three curves, $\mu _{r}^{o}(p),$ $\mu _{r}^{e}(p)$ and $\mu 
_{r}(p),$corresponding to the odd, even and total number of appearances 
of the histories, respectively. It is easy to distinguish three regions: 
($i$) zone 1, where $\mu _{r}^{o}(p)\approx 0.25\ll \mu _{r}^{e}(p).$ 
This is the PTD region, as can be verified by the simple observation of 
the record of minorities. As it is known, 0.25 is the expected value for 
a random set; ($ ii$) zone 2, an intermediate region where $\mu 
_{r}^{o}(p)$ is no longer that of a random process, but still $\mu 
_{r}^{o}(p)<\mu _{r}^{e}(p);$ and finally, ($iii$) zone 3, a region 
where the even and odd curves coincide and, therefore, there is no PTD 
present. Notice that the width of the intermediate zone shrinks as $N$ 
increases. For $m=2$ and $N=3721,$ the 3rd zone starts at $\ p^{\ast 
}\simeq 0.5$ (this value has only a moderate size effect).

The behavior for small values of $p$ can be understood as a simple 
extension of the behavior of the MG:\ there is only a small number of 
$I.A.$, so that they can reduce somewhat the value of $\sigma$, but can 
not change the general outcome of the game. It is also possible to 
explain in simple terms the behavior for $p=1$. In this case, $all$ the 
players are interacting, hence each one has a neighborhood of 4 $I.A.$ 
Let us assume, for the sake of concreteness, that the minority outcome 
for an odd appearance of a given history is $\mathcal{M}=1$; we could 
expect that in the next (even)\ appearance of the same history, the 
majority of the players will pick the same side, as the PTD indicates. 
With the new rules, however, there is an extra opportunity for all the 
players. Now, most of the agents will notice that their neighbors will 
chose `1', and therefore each one in this condition will, in turn, 
change her bet. As a result, the majority will chose `0', and the 
minority will again be `1'; virtual points are given to the $same$ group 
of strategies than before, in agreement with the previous crowd effect: 
each history will be followed always by the same answer. One consequence 
of this behavior, is that the system will fall into a cyclic evolution, 
with a period equal to $4$ (the number of histories), or smaller.

Notice that, in general, $p^{\ast }$ is a function of $z$, 
$p^{\ast}(z)$. In the region between $p^{\ast }$ and $p=1$, the system 
behaves qualitatively in the same form. We found it appropriate to 
analyze the sequence of minorities of the game, \{$\mathcal{M}_{i}$\}, \ 
determining the \emph{entropy rate }\cite{Cover+Thomas} of the set. The 
entropy rate of $n$ random variables \{$X_{i}$\} measures how does the 
entropy of the sequence growth with $n$. If the stochastic process is a 
stationary Markov chain, the entropy rate $H_{T}(X)$ can be written as 
follows: \ 

\begin{equation}
H_{T}(X)=-\sum_{i,j}\lambda _{i}P_{ij}\log _{2}P_{ij}  \label{entropy rate}
\end{equation}

In this equation, $\lambda =(\lambda _{1},\lambda _{2},...)$ is a vector 
whose components are the stationary probabilities of \ $X_{1}$, $X_{2}$, 
etc. $P$ is a probability transition matrix, whose elements measure the 
probability of the transition between the states $i$ and $j,$ $i.e$. $ 
P_{ij}=prob\{X_{n+1}=j|X_{n}=i\}.$ The reason to consider here this 
magnitude is the following:\ if the sequence of $m$-histories resulting 
from \{$\mathcal{M}_{i}$\} has a cyclic behavior, the corresponding 
entropy rate is zero, because in this case $P_{ij}$ is either 0 or 1. In 
other words, $ H_{T}$ \emph{is an efficient detector of cyclic 
behavior}. From a physical point of view, once one knows in what kind of 
cycle is the system, there is no uncertainty about the next time step; 
that is the reason why the entropy rate vanishes.

\begin{figure}[tbp]
\includegraphics[width=12cm,clip]{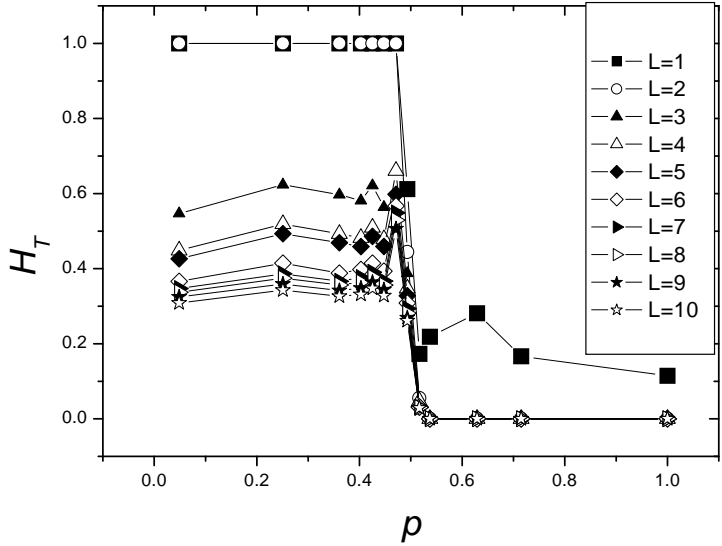}
\caption{Entropy rate $H_T$ as a function of $p$, for $N=3721$ agents 
and $m=2$. The symbols representing the different word sizes are shown 
in the inset. In this case $p^{\ast} \simeq 0.5$. Notice that for 
$p<p^{\ast}$, and $L \leq m$ the entropy rate is $H_T=1$, while for $p > 
p^{\ast}$ and $L \geq m$ the entropy rate is $H_T=0$.}
\label{HT}
\end{figure}

The set \{$\mathcal{M}_{i}$\} can be studied by forming $M-L+1$\ `words' 
of length $L$ out of the sequence of $M$ minority sides; two successive 
words will share $L-1$ values, or bits; thus for example if we want to 
look at words of length $L=3$, and the sequence of minorities is 
represented by $...abcde...$ (where each letter can take the values 
${0,1}$) one word will be `abc' and the following `bcd', etc. 
\cite{palabras}.  We determined $H_{T}$ for $ m$ = 2, 5  and 8, and 
several values of $p$ in the interval \{0,1\}, using word sizes $L$ = 1 
to 10 (see Fig.\ref{HT}). For $ p\gtrsim p^{\ast }$ we found that 
$H_{T}\approx 0,$ indicating the presence of a cyclic evolution. In 
fact, a close inspection of the series of minority sides shows that for 
$p$ in the interval \{$p^{\ast },1\},$ the system falls in a cyclic 
motion for a rather long number of time steps, until it suddenly 
switches to another cycle, and so on. The contribution to the entropy 
rate from the transition time steps between two different cycles is very 
small, but noticeable. As $p$ approaches $p=1,$ the change of cycle 
becomes more and more uncommon.

Two other aspects of the results shown in Fig.\ref{HT} deserve to be 
emphasized. In the first place, it is rather clear that the general 
behavior of the entropy rate follows that of the reduced variance, in 
the sense that both show two different `phases', for the same values of 
$p$: one `phase' for $p\leq p^{\ast },$ and a different one for 
$p>p^{\ast }.$ On the other hand, the entropy rate is roughly constant 
in the $p$-region where the system follows the PTD. Moreover, the 
entropy rate for $L\leq m$ takes the value $H_{T}\simeq 1$, showing that 
for these word sizes the minority set is essentially random. Indeed, it 
follows from the MG rules that each step $i$ of the game can only be 
followed by one of the two possible outcomes (\emph{i.e.} either a zero 
or a one); if the behavior is random, it turns out that $P_{ij}$ can 
only be equal to zero (if $i$ and $j$ can not be connected by the MG 
rules) or 1/2. On the other hand the summation over the $\lambda$ values 
in Eq.\ref{entropy rate} will be equal to one; therefore it follows that 
$H_{T}=1$ for a random set built with the MG rules. Following the 
description of Manuca $et.al$ \cite{Manuca}, it can be said that one can 
not get information `hidden' in \{$\mathcal{M}_{i}$\} by using words of 
length $L \leq m$: this will only be possible if one uses words one bit 
(or more) longer than $m$.  

\begin{figure}[tbp]
\begin{tabular}{cc}
\includegraphics[width=6.5cm,clip]{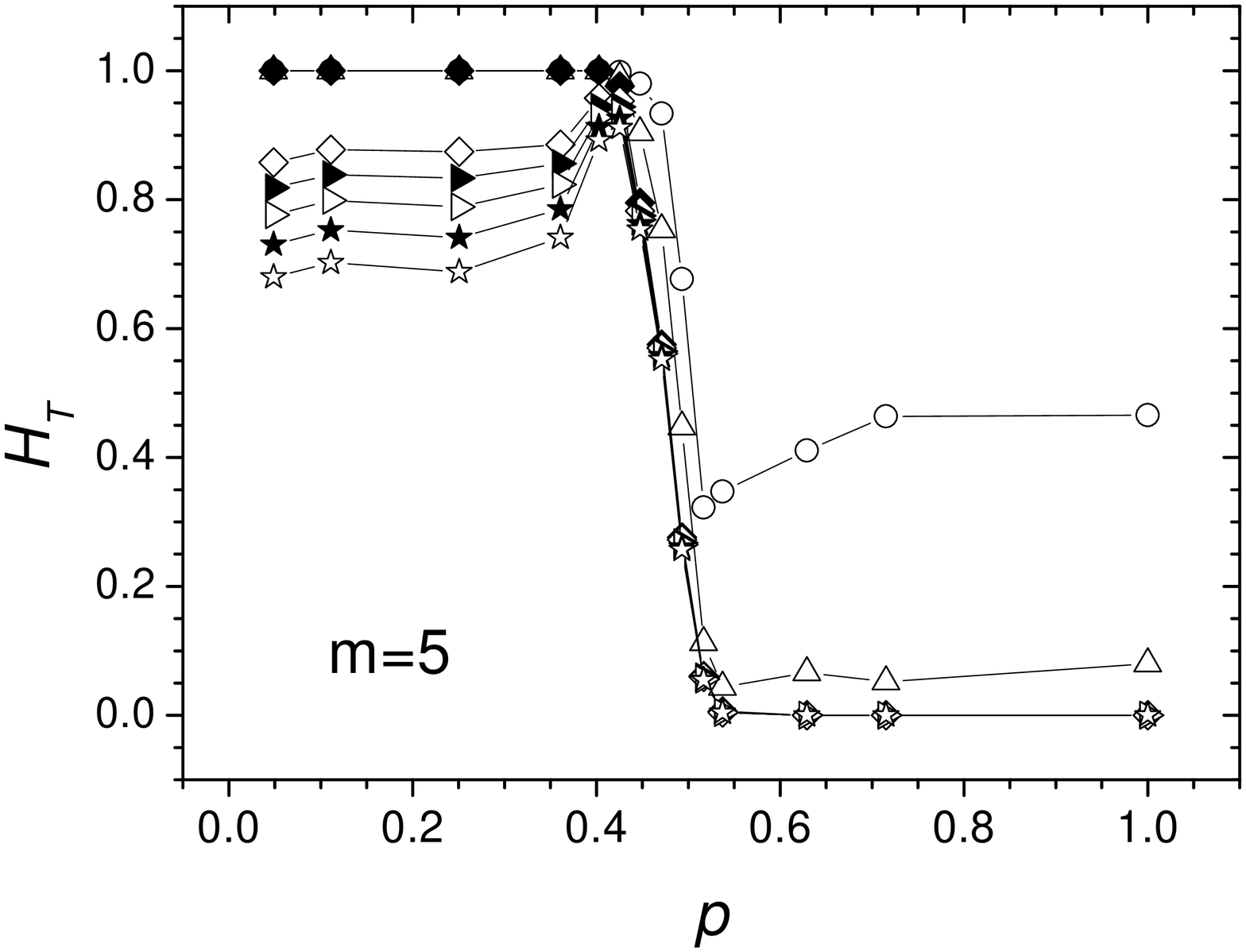} & \includegraphics
[width=6.5cm,clip]{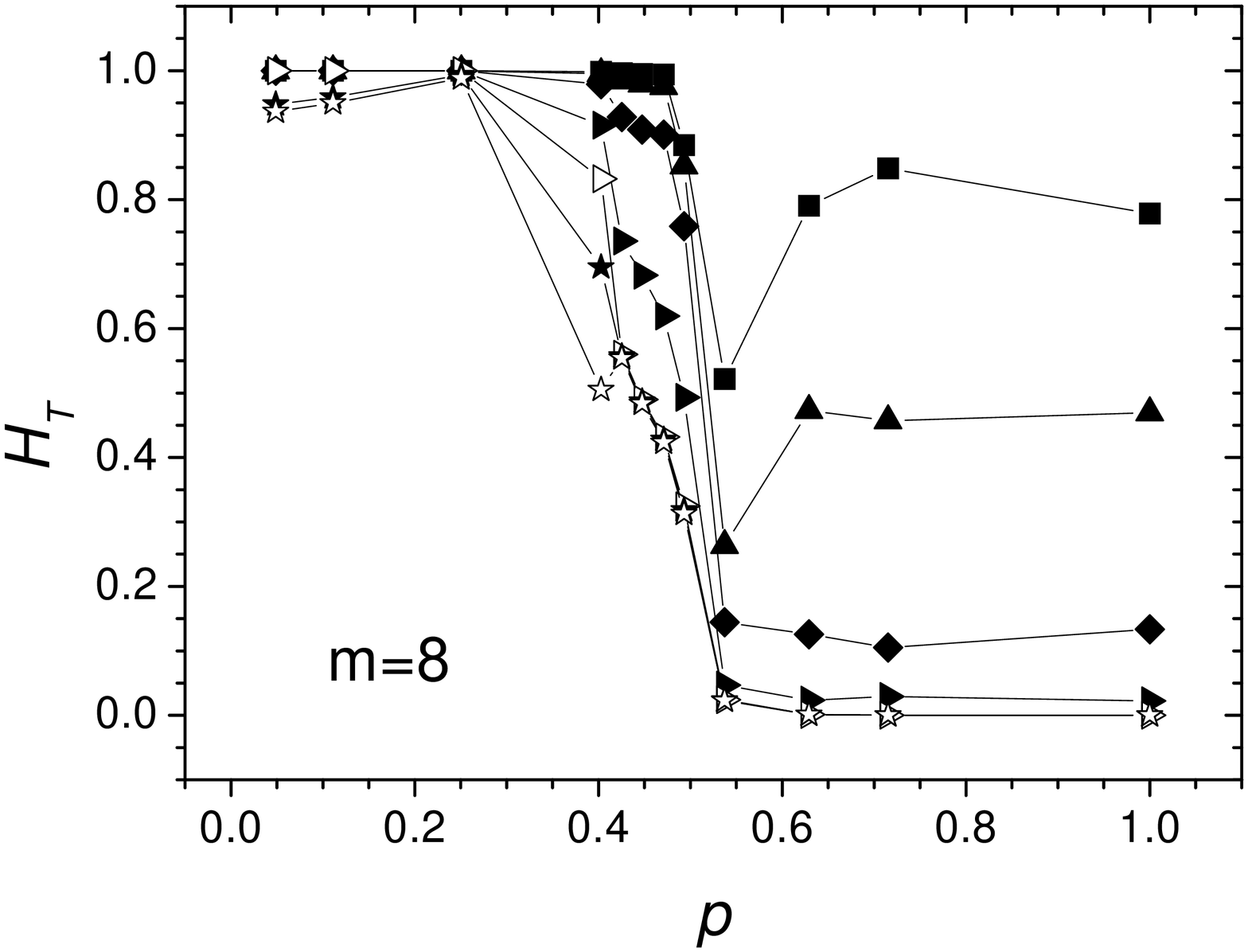} \\
\end{tabular}
\caption{Entropy rate as a function of $p$, for $m$ = 5 and 8. We use 
the same symbols as in Fig.\ref{HT}. For clarity we only show results 
for selected values of $L$.}
\label{otros_m}
\end{figure}

Figure \ref{otros_m} shows results for $m=5$ and $8$. It can be seen that 
they are similar to those described above for $m=2$, the main difference 
being that the value of $p^{\ast }$ gets smaller as $m$ increases. 
Notice that for $N$=3721, the case with $m$=8 has $z \simeq$ 0.07, 
$i.e.$ it still belongs to the first region. 

\section{Conclusions \label{conclusiones}}

We have studied a modification of the rules of the Minority Game, hoping 
to retain the main ideas of the model, and at the same time introducing 
a more realistic description of the agents. This was done through the 
introduction of certain number of interacting agents, that can know the 
decisions of their nearest neighbors, before they make their choice. As 
a result, the efficiency of the system improves if the fraction $p$ of 
$I.A.$ is small, but it is worst when this fraction approaches $p\simeq 
1.$ The relation of the reduced variance $\mu _{r}$ with $p$, for 
$z<z_c$ resembles the relation of $\mu _{r}$ with $m$; in fact, for 
every value of $z$ there is an optimal value of $p$, $p^{\ast }(z)$, for 
which one has the minimum value of $\mu_r$. Thus, for a small fraction 
of $\ I.A.$ the system remains in the PTD regime, sweeping through the 
whole set of histories; but when this fraction is greater than $p^{\ast 
}(z)$ (but still within the first, informationally efficient region), it 
falls into a cyclic, and in general non ergodic motion. In this sense, 
the use of the entropy rate proved to be a sensible tool to study the 
sequence of histories visited.


\begin{thebibliography}{99}
\bibitem{palabras} We have also studied a related set, by choosing words 
of length $L$ that are {\it contiguous}, but do not share any bit (one 
can form $M/L$ words of this kind). We found that the periodic behavior 
is also detected by this set of words, but its behavior for small values 
of $p$ is not very clear.
\end{thebibliography}
\end{document}